\documentclass[useAMS,usenatbib]{mn2e}
 
\usepackage{amsmath}
\usepackage{float}
\usepackage{graphicx}
\usepackage{txfonts}
\usepackage{indentfirst}
\usepackage{color}
\usepackage{ulem}
\usepackage{hyperref}

\newcommand{\eq}[1]{\begin{equation}  #1 \end{equation}}

\newcommand{\dd}{{\rm d}}

\def\apj{ApJ}

\def\apjs{ApJS}
\def\aap{A\&A}

\def\mnras{MNRAS}

\title[Comparison with hydrodynamics simulation]{Analytical model
for non-thermal pressure in galaxy clusters - II. Comparison with cosmological
hydrodynamics simulation}

\author[Xun Shi, Eiichiro Komatsu, Kaylea Nelson and
Daisuke Nagai]{Xun Shi$^{1}$\thanks{E-mail:
xun@mpa-garching.mpg.de}, Eiichiro Komatsu$^{1,2}$,
Kaylea Nelson$^{3,5}$,
Daisuke Nagai$^{3,4,5}$ \\
$^{1}$Max-Planck-Institut f\"ur Astrophysik,
Karl-Schwarzschild-Stra{\ss}e 1, D-85740 Garching bei M\"unchen, Germany\\
$^{2}$Kavli Institute for the Physics and
Mathematics of the Universe (Kavli IPMU, WPI), Todai Institutes for Advanced Study, the
University of Tokyo,\\ Kashiwa 277-8583, Japan\\ 
$^{3}$Department of Astronomy, Yale University, New Haven, CT 06520, U.S.A.\\
$^{4}$Department of Physics, Yale University, New Haven, CT 06520, U.S.A.\\
$^{5}$Yale Center for Astronomy \& Astrophysics, Yale University, New Haven, CT
06520, U.S.A.
}

\begin{document}

\pagerange{\pageref{firstpage}--\pageref{lastpage}} 

\maketitle

\label{firstpage}

\begin{abstract}
Turbulent gas motion inside galaxy clusters provides a non-negligible
 non-thermal pressure support to the intracluster gas. If not corrected,
 it leads to a systematic bias in the estimation of
 cluster masses from X-ray and Sunyaev-Zel'dovich (SZ)
 observations assuming hydrostatic equilibrium, and affects
 interpretation of measurements of the SZ power spectrum and
 observations of cluster outskirts from ongoing and upcoming large
 cluster surveys. Recently, \citet{shi14} developed an analytical model for
 predicting the radius, mass, and redshift dependence of
 the non-thermal pressure contributed by the kinetic random motions
 of intracluster gas sourced by the cluster mass growth.
 In this paper, we compare the predictions of this analytical model to a
 state-of-the-art cosmological hydrodynamics simulation. As different mass
 growth histories result in different non-thermal pressure, we perform
 the comparison on 65 simulated galaxy clusters on a cluster-by-cluster basis.
 We find an excellent agreement between the modelled and simulated non-thermal
 pressure profiles. Our results open up the possibility of using the
 analytical model to  correct the systematic bias in the mass estimation
 of galaxy clusters. We also  discuss tests of the physical picture
 underlying the evolution of intracluster non-thermal gas
 motions, as well as a way to further improve the analytical modeling, which may help achieve a unified understanding of non-thermal phenomena in galaxy clusters.
\end{abstract}


%%%%%%%%%%%%%%%%%%%%%%      sec       %%%%%%%%%%%%%%%%%%%%%%%%%
\section[]{Introduction}
Precise mass determinations of galaxy clusters are crucial for their
cosmological applications. We usually assume hydrostatic
equilibrium between the pressure gradient and the gravitational force on
the intracluster gas when determining masses from
X-ray and Sunyaev–Zel'dovich (SZ) observations. These observations, however, typically measure only the 
thermal pressure of the gas. Non-thermal pressure, if neglected, introduces a
bias in the hydrostatic mass estimation (HSE mass bias). This would, in turn,
bias the cosmological constraint from the cluster mass function and the SZ power
spectrum, and affect the interpretation of observations of
cluster outskirts from ongoing and upcoming large cluster surveys.

Observationally, the HSE mass bias manifests itself
 as a systematic difference between the X-ray (or SZ) derived mass and
the lensing mass of up to $30\%$ (\citealt[][]{allen98, mah08, richard10,
zhang10, vdl14}, but see also non-detections, e.g., \citealt{israel14}).
Hydrodynamics numerical simulations of intracluster gas using both grid-based
\citep[][]{iapichino08,vazza09,lau09,iapichino11,nelson14,nelson14b} and
particle-based \citep[][]{dolag05,vazza06,bat12} methods have found that 
 the intracluster gas motions generated in the structure formation
 process contributes significantly to the non-thermal pressure. These alone lead to an HSE mass bias
comparable to that found from observations \citep[][]{rasia06,rasia12,nagai07b,piff08,lau09,men10,nelson12}.
In addition to the structure formation process, turbulent gas motions can be
generated in the cluster outskirts by the magnetothermal instability
 \citep{par12, mcc13}, and in the cluster core by core sloshing and energy
 injection from black holes and stars. Magnetic fields and cosmic rays can also
 potentially contribute to the non-thermal pressure. Residual acceleration of
 gas, apart from the non-thermal pressure, introduces an additional
 source of deviation from the hydrostatic equilibrium
 \citep{lau13,suto13,nelson14}.
We refer to, e.g., \citet{shi14} for a discussion of these sources, and
focus, in the following, on the pressure support from the
intracluster gas motions generated during structure formation.

Due to the high Reynold number associated with typical intracluster
gas motions, the intracluster gas flow is highly turbulent. Nevertheless, since
the turbulence cascade time-scale on galaxy-cluster-size scales can be comparable to the Hubble time, the
 existence of large scale coherent motions is expected. Current
 hydrodynamic simulations cannot yet achieve the high Reynold number
 characteristic of true turbulence. Still, according to the physical scale of
 the resolved motions, it is possible to distinguish motions that appear
 random or coherent on a certain scale and refer to them as 
 `turbulence' and `bulk motions', respectively. When estimating the
 non-thermal pressure and the HSE mass bias, however, it is not necessary to
 distinguish `turbulence' and `bulk motions', as both of them contribute in the
 same way \citep{lau13}. In the following, we follow \citet{shi14} and refer to
 the non-thermal random motions in the diffuse intracluster gas as `turbulence' or `turbulent gas
motions'. Note that the gas motions associated to
self-gravitating substructures, on the other hand, do not contribute to the HSE mass bias in general. 

Several observations have provided indirect evidence for the
intracluster gas motions: measurements of
the magnetic field fluctuations in diffuse cluster radio sources
\citep[][]{mur04, vogt05, bon10,vacca10}, X-ray surface
brightness fluctuations or pressure fluctuations inferred from X-ray maps
\citep{schuecker04, churazov12, sim12}, and the non-detection of resonant
scattering effects in the X-ray spectra \citep{churazov04}. Future observations
of the X-ray emission lines are considered as the most promising
method to measure turbulence velocities directly
\citep[][]{sunyaev03,zhur12,shang12}, and so far the method has
already provided a few upper limits in the cluster cores
\citep{sanders11,sanders13}. Whereas these observations greatly contribute to
our understanding of the non-thermal phenomena in the intracluster gas, it is hard to use them to
estimate the turbulence pressure accurately. Moreover,
these observations are mostly limited to nearby clusters or the inner
regions with high surface brightness (see \citealp{nagai13} for an estimation of the detectability of intracluster gas motions by the upcoming \textsl{Astro-H} mission).

On the other hand, the mass estimates require an accurate determination of
the non-thermal pressure in the outskirts of clusters where most of the
mass resides.  
Therefore, the amplitude of intracluster turbulence pressure
in the outskirts has to be derived
theoretically from the existing knowledge of the injection and dissipation of
intracluster turbulence.  

One way to estimate the turbulence pressure is to measure it from cosmological
hydrodynamics simulations.
However, since a large, high-precision light-cone hydrodynamics simulation is
still too expensive to carry out, it is desirable to have an analytical model
that can predict the turbulence pressure, either alone or combined with 
dark matter only N-body simulations. More importantly, an analytical model is
based on physical understandings. Thus, by comparing the predictions drawn from
an analytical model to simulations and observations, the physical
understandings can be tested and improved, forming a healthy feedback loop.

To this end, \citet[][hereafter SK14]{shi14} developed an
analytical model for computing the time evolution of the intracluster turbulence
pressure. The model is based on a physical picture of turbulence injection
during hierarchical cluster mass assembly, and turbulence dissipation with a time-scale
determined by the turnover time of the largest turbulence eddies. In
this paper, we shall compare the turbulence pressure predicted by this
analytical model to that measured in a state-of-art cosmological
hydrodynamics simulation. This comparison will test the validity of the
analytical model as well as some aspects of the underlying physical picture.

The rest of the paper is organized as follows.
In Sect.\;\ref{sec:sim}, we introduce the simulation and the cluster sample used
for the comparison.
In Sect.\;\ref{sec:method}, we demonstrate how to
apply the analytical model to the simulation data.
In Sect.\;\ref{sec:nonth}, we present and discuss the results.  
The underlying physical picture of turbulence injection
and dissipation, as well as how to test them more thoroughly, are discussed in
Sect.\;\ref{sec:dis}.
We conclude in Sect.\;\ref{sec:con}.

%%%%%%%%%%%%%%%%%%%%%%      sec       %%%%%%%%%%%%%%%%%%%%%%%%%
\section[]{Simulation and cluster sample}
\label{sec:sim}

\begin{figure}
\centering
\includegraphics[width=0.4\textwidth]{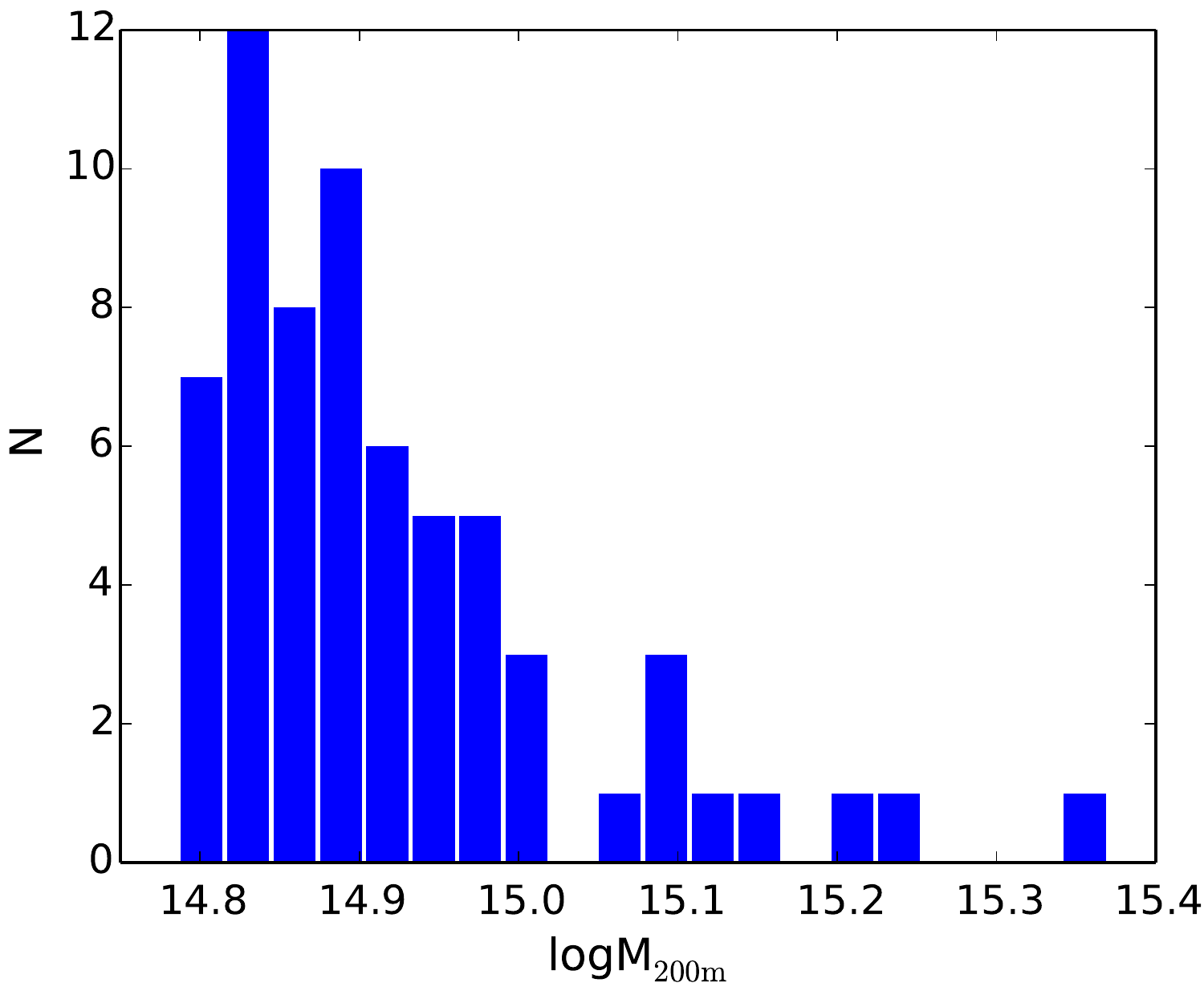}
\caption{Distribution of cluster masses at $z=0$ in the mass-limited
 sample of simulated galaxy clusters.}
\label{fig:0}
\end{figure}
    
\begin{figure}
\centering
\includegraphics[width=0.41\textwidth]{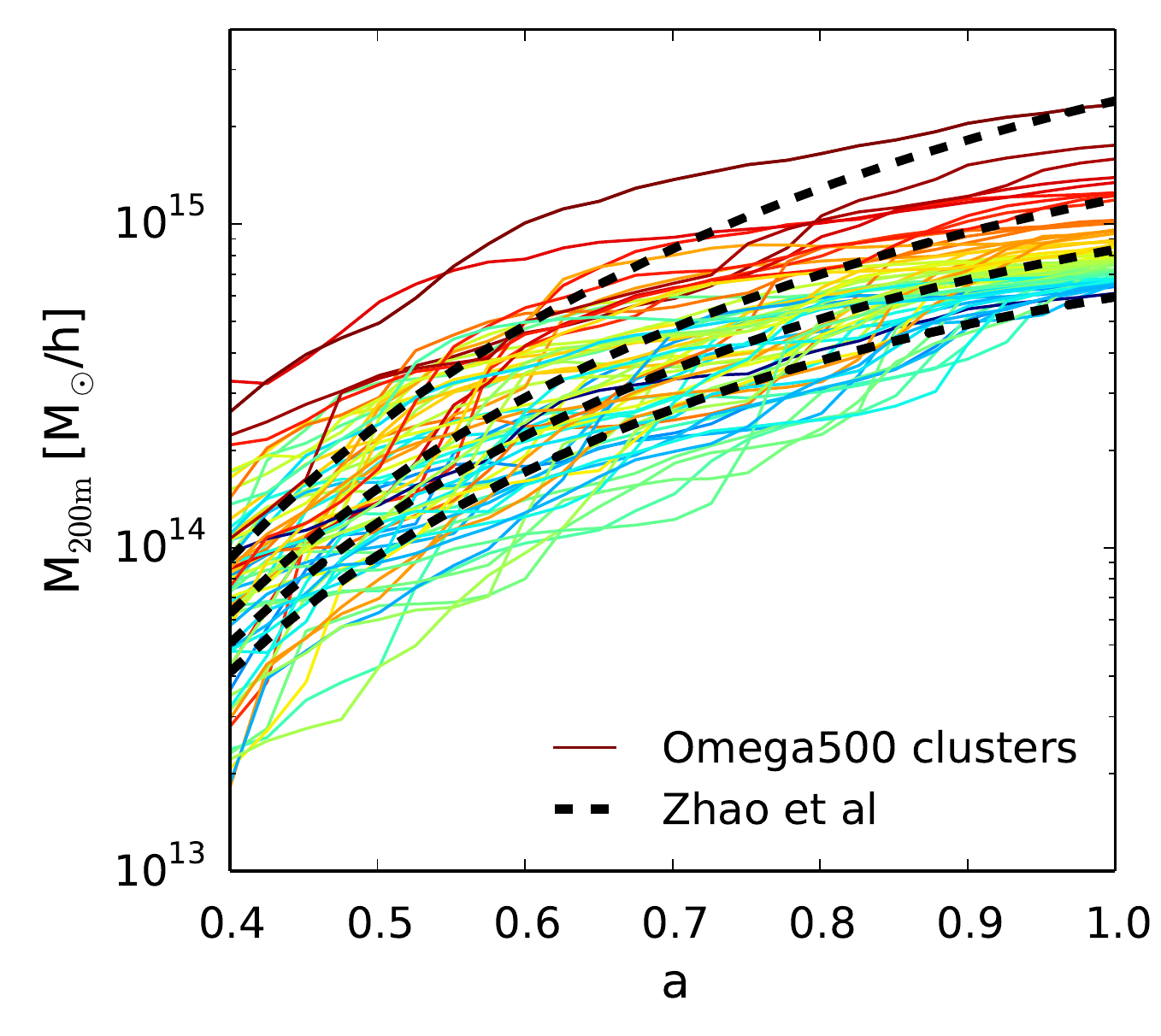}
\caption{Mass accretion histories of the mass-limited sample of 65 clusters
from the Omega500 simulation. Each solid line shows the mass accretion
history of one simulated cluster, colour-coded according to its mass at $z=0$
($a=1$ where $a$ is the scale factor). We also show the mean halo mass
accretion histories of four different halo masses computed using the
\citet{zhao09} method (black dashed lines).}
\label{fig:mah}
\end{figure}

We compare the SK14 model with the outputs of the Omega500
simulation \citep{nelson14},  
a large cosmological Eulerian simulation 
 performed with the Adaptive Refinement Tree (ART)
N-body+gas-dynamics code 
\citep{krav99,krav02,rudd08}. In order to
achieve the dynamic ranges necessary to resolve the cores of haloes, adaptive
refinement in space and time and non-adaptive refinement in mass
\citep{klypin01} are used. The simulation has a comoving box length of $500$ h$^{-1}$Mpc and a maximum comoving spatial resolution of $3.8$
h$^{-1}$kpc, and is performed in a flat $\Lambda$CDM model with the \textsl{WMAP}
five-year cosmological parameters \citep{kom09}.
For consistency with the physics included in the analytical model, the
simulation we use does not include radiative cooling or feedback. See
\citet{nelson14b} for the implications of neglecting these additional
physics in simulations.

We select
a mass-limited sample of $65$ galaxy clusters at $z=0$ from the
simulation. Its mass distribution is shown in Fig.\;\ref{fig:0}. 
We measure
one-dimensional profiles of various quantities such
as the density and pressure at $25$ snapshots between $z=0$
and $z=1.5$. See \citet{nelson14} and \citet[][hereafter NLN14]{nelson14b} for
more information on the simulation and the cluster sample.

Fig.\;\ref{fig:mah} shows the mass accretion histories of the cluster
sample. Each of the 65 clusters is assigned a colour depending on
their final mass at $z=0$. The mass, $M_{\rm 200m}$, is defined as
the mass enclosed within 
the radius, $r_{\rm 200m}$, within which the average matter density
equals $200$ times the mean mass density of the universe.
The dashed lines in Fig.\;\ref{fig:mah}  show the
analytical mean halo mass accretion histories of \citet{zhao09} for four
representative halo masses.   
We find that the mass accretion histories of the simulated clusters largely
agree with that predicted by \citet{zhao09}, despite that the most massive
clusters in the sample show slightly slower mass accretion histories
than the prediction of \citet{zhao09}. This suggests that the few most
massive clusters in the simulated cluster sample can be slightly more
relaxed than the cosmic average. We do not expect this to affect 
generality of our results.

%%%%%%%%%%%%%%%%%%%%%%      sec       %%%%%%%%%%%%%%%%%%%%%%%%%
\section[]{Analytical model of non-thermal pressure}
\label{sec:method}

\subsection{The model}
The SK14 model uses a first-order differential equation
\eq{
\label{eq:sigkin}
\frac{\dd \sigma^2_{\rm nth}}{\dd t} = -\frac{\sigma^2_{\rm nth}}{t_{\rm d}}
+ \eta\;\frac{\dd \sigma^2_{\rm tot}}{\dd t}\,,  
}
 to describe the time evolution of
turbulence velocity dispersion squared, $\sigma^2_{\rm nth}$, which is also the
turbulence pressure $P_{\rm nth}$ per unit density, i.e.,
$\sigma^2_{\rm nth} \equiv P_{\rm nth}/\rho_{\rm gas}$.
The evolution of $\sigma^2_{\rm nth}$ is sourced by that of
the total velocity dispersion squared,
$\sigma^2_{\rm
tot}$, which is the sum of turbulence (`nth', non-thermal) and thermal (`th')
velocity dispersion squared: 
\eq{
\label{eq:sigtot}
\sigma^2_{\rm tot} \equiv \frac{P_{\rm th}}{\rho_{\rm gas}} + \sigma^2_{\rm nth}
= \frac{P_{\rm tot}}{\rho_{\rm gas}}\,,} 
with $P_{\rm tot} \equiv P_{\rm th} + P_{\rm nth}$.
The turbulence dissipation time-scale, $t_{\rm d}$, is taken to be
proportional to the dynamical time of the intracluster gas, $t_{\rm d} =
\beta t_{\rm dyn}/2$. It can be derived from the accumulated total mass
profile, $M(<r)$, as the dynamical time is defined by $t_{\rm dyn}\equiv
2\pi\sqrt{r^3/[GM(<r)]}$. 
In general, $\sigma^2_{\rm tot}$, $t_{\rm d}$, and hence $\sigma^2_{\rm nth}$,   
are all functions of radius, mass, and redshift of a cluster. The two parameters
in the model, $\eta$ and $\beta$, are taken to be constants by assumption.

We need $\sigma^2_{\rm tot}$ and $t_{\rm d}$ to solve
equation~(\ref{eq:sigkin}). These quantities are, to first order,
dictated by the gravitational potential. The essential input knowledge
is then how the gravitational potential deepens with time, or simply the
mass accretion history. Different clusters have different mass accretion
histories; thus, to compare the model predictions with the simulated clusters
on a cluster-by-cluster basis, we take  $\sigma^2_{\rm tot}$ and
$t_{\rm d}$ directly from the simulation outputs of individual
clusters.

We measure the turbulence velocity dispersion, $\sigma_{\rm
nth}$, in each radial shell as the rms velocity after subtracting the
mean velocity of the shell with respect to the center-of-mass velocity
of the total mass interior to this radial shell (NLN14).  In
Appendix.\;\ref{sec:ap1} we will explain and discuss our procedure of measuring
$\sigma_{\rm nth}$ in detail. We then compute the total velocity dispersion squared, $\sigma^2_{\rm tot}$, according to equation~(\ref{eq:sigtot}) \footnote{{Alternatively, one
may compute} $\sigma^2_{\rm tot}$ from $P_{\rm tot}$ which by itself is computed
using the hydrostatic equilibrium equation, and then derive 
$\sigma^2_{\rm nth}$ as the difference of $\sigma^2_{\rm tot}$ and
$P_{\rm th}/\rho_{\rm gas}$. Since simulated galaxy
clusters are not spherically symmetric nor fully relaxed, this
alternative method yields slightly different $\sigma^2_{\rm tot}$.
While the $\sigma^2_{\rm tot}$ profiles of the cluster sample
computed with the two methods are very similar in the virial region of
the clusters, the $\sigma^2_{\rm nth}$ profiles are
significantly different because the alternative method computes
$\sigma^2_{\rm nth}$ as the difference of two large quantities. Since
 $\sigma^2_{\rm nth}$ 
computed this way is more prone to numerical errors, we choose
the method described in the main text to compute $\sigma^2_{\rm nth}$
and $\sigma^2_{\rm tot}$.} , and
compute the non-thermal pressure fraction, $f_{\rm nth}$, as
their ratio, i.e., $f_{\rm nth} \equiv {\sigma^2_{\rm nth}}/{\sigma^2_{\rm tot}}$.
 Since $f_{\rm nth}$ is typically much smaller than unity,
$\sigma^2_{\rm tot}$ is mainly contributed by the thermal velocity dispersion
and can be regarded, to the first order, as being proportional to the gas
temperature.

\subsection{Smoothing the source term}

\begin{figure}
\centering
\includegraphics[width=0.45\textwidth]{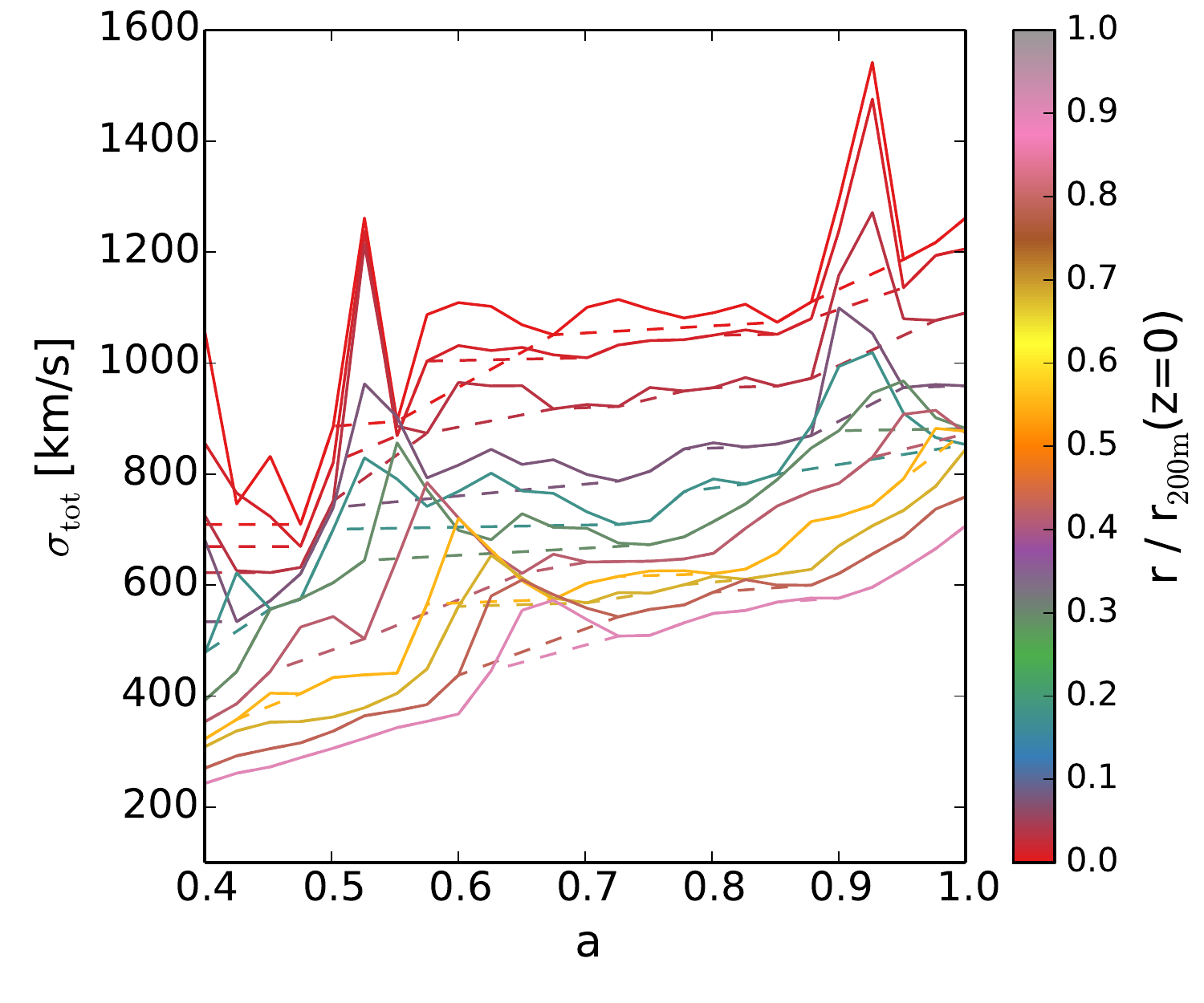}
\caption{Growth of $\sigma_{\rm tot}$ as a function
of the scale factor of the universe in one representative cluster with a typical
mass and accretion history.
Each solid line shows $\sigma_{\rm tot}$ measured in the simulation at a
 certain Eulerian radius indicated by the colour bar. The dashed lines
 are the smoothed $\sigma_{\rm tot}$ growth curves used in
 the modeling.} 
\label{fig:02}
\end{figure}

As a cluster grows in mass, its $\sigma_{\rm tot}$ generally increases,
suggesting a positive source term in the right hand side of
equation~(\ref{eq:sigkin}). For the 
simulated clusters, however, the $\sigma_{\rm tot}$ at each Eulerian radius may
also decrease due to local inhomogeneities. 
As an example, in Fig.\;\ref{fig:02} we show $\sigma_{\rm tot}$ at a few
radial bins of one cluster as a function of the scale
factor of the universe. The selected cluster has a mass and an
accretion history both close to the median of the mass-limited cluster sample. 
Some wiggles exist in $\sigma_{\rm tot}$, which propagate from small to large
radii. They likely share the same origin with the peaks in the
hydrostatic mass estimates in fig.\;2 of \citet{nelson12}, and correspond to outwardly
moving merger shocks that sweep across the cluster region in a time of 1--2~Gyr
(see Appendix.\;\ref{sec:ap2}).

The analytical model does not intend to capture such transient phenomena,
but rather their long-term effect on the intracluster medium. Therefore we
smooth these wiggles to reduce their numerical effect. We
do so by choosing the points from each simulated $\sigma_{\rm tot} (a)$
curve which have a smaller value than all the points to the right of
this curve (at a larger $a$), and fit linearly between these chosen
points. We then use the
resulting monotonously increasing $\sigma_{\rm tot} (a)$ (dashed lines in
Fig.\;\ref{fig:02}) in the modeling. We note that this smoothing
is only performed for the source term in the analytical model, but not in any
simulation data used for the comparison. Radial bins that happen to be at
the disturbed state (corresponding to the peak of the wiggles) at the time of
the comparison would explain part of the scatter in the comparison.

\subsection{Initial condition} 
In SK14 we have argued that, as long as the initial time is chosen to be early
enough, the choice of the value of $f_{\rm nth}$ at the initial time
does not affect the final value of $f_{\rm nth}$. In the inner region of
the cluster, this is because the short turbulence dissipation time
drives $f_{\rm nth}$ quickly to its limiting value determined by the
ratio of $t_{\rm d}$ and the cluster mass growth time-scale (see
Sect.\;3.2 of SK14). In the cluster outskirts,  
the turbulence pressure does accumulate throughout time, but the growth is
significant after the region enters the virial radius of the cluster, which
occurs only at late times. 

In this paper, the initial time is chosen at $z=1.5$,
which is early enough for the above arguments to hold to a high degree
of accuracy for studying cluster profiles at $z=0$. Thus, for
convenience and consistency, we choose the initial condition to be $f_{\rm nth}
= 0$ for all clusters at $z=1.5$. Another option, namely using the values of
$f_{\rm nth}$ measured from the simulation at $z=1.5$, can provide a more
precise initial condition, but only for regions inside clusters which are
dynamically relaxed at that time. We have compared the $f_{\rm nth}$ values at
$z=0$ using this initial condition with those using the default initial
condition. The difference is negligible inside $r_{\rm 200m}$.

%%%%%%%%%%%%%%%%%%%%%%      sec       %%%%%%%%%%%%%%%%%%%%%%%%%
\section[]{Results: model versus simulation}
\label{sec:nonth}

We shall limit the comparison between the model predictions
and the simulation outputs to $(0.1 - 1) r_{\rm 200m}$. We restrict the study
to $r<r_{\rm 200m}$ (about $1.3 r_{\rm vir}$ at $z=0$ and $r_{\rm vir}$ at $z=1$ in a standard $\Lambda$CDM
cosmology for cluster-mass objects) to avoid the region where there is
significant gas infall. 
We also avoid the cluster core region ($r<0.1r_{\rm 200m}$) because of both
theoretical and numerical difficulties there, such as the uncertainty on the
feedback effect of the central AGNs, the disagreement of numerical
methods on gas thermodynamical quantities in the core region, and the ambiguity in the
choice of the cluster center and its consequence on the projected
one-dimensional profiles. Outside the core region the
observational measurement of the velocity field becomes exceedingly difficult,
implying that the observational test of the model may be restricted to
a few nearby systems even with upcoming instruments. This, on the other hand,
emphasizes the importance of analytical understanding of the non-thermal
pressure, as well as the comparison of analytical models and numerical
simulations. 

We choose $\beta=1$ and $\eta=0.7$ as the preferred value
(SK14). Effects of varying $\beta$ and $\eta$ will be presented in
Sect.\;\ref{sec:varyparam}. All comparison will be performed on the cluster
sample at $z=0$.

\subsection{Non-thermal pressure fraction}

\begin{figure}
\centering
\includegraphics[width=0.45\textwidth]{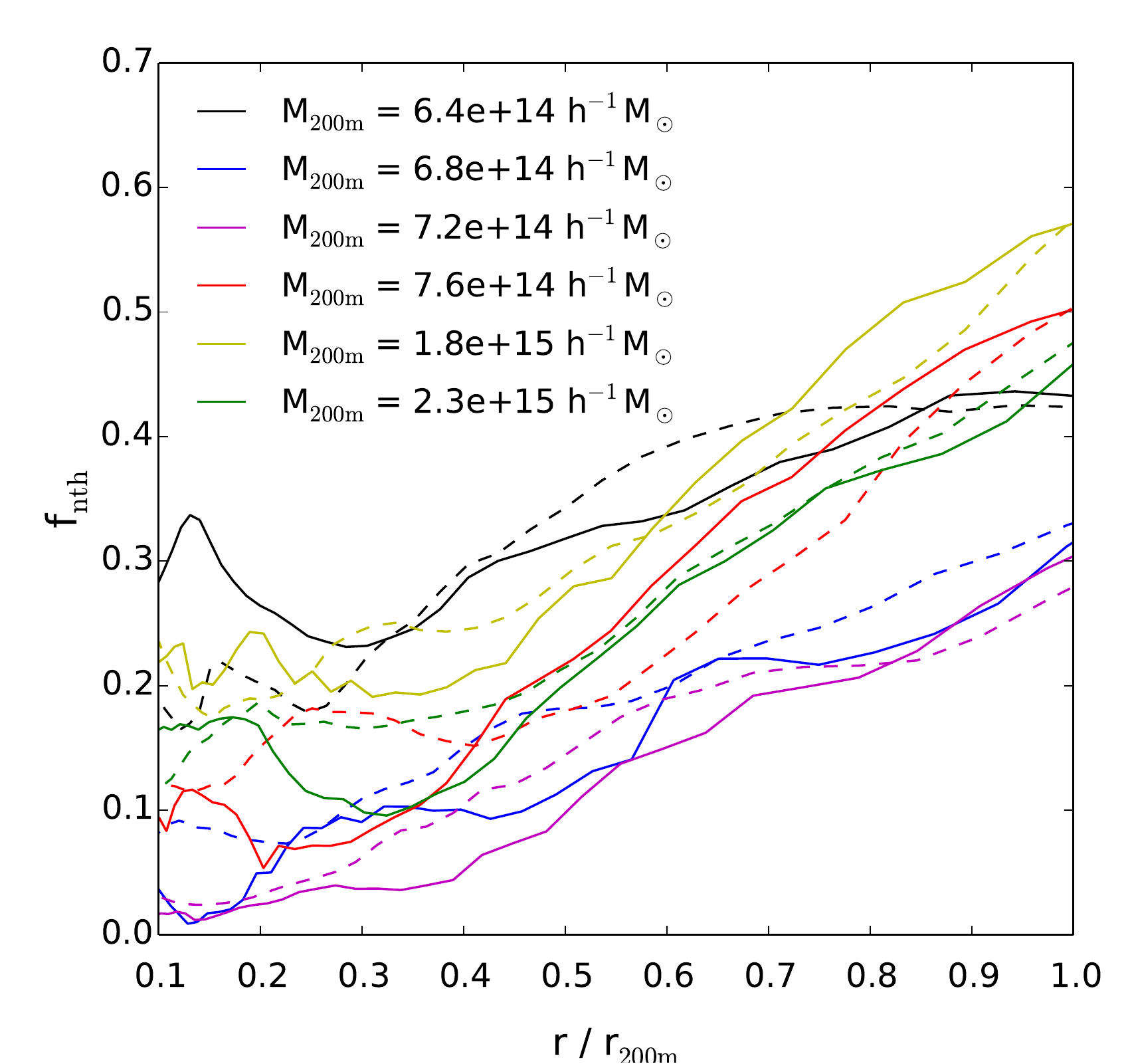}
\caption{Comparison of modelled (solid lines) and simulated (dashed lines)
$f_{\rm nth}$ profiles of individual clusters. Profiles of six typical 
clusters with a spectrum of different masses at $z=0$ are shown. 
The radius is scaled with $r_{\rm 200m}$, which has a value of about 2.3 Mpc/h
for a cluster with the median mass of the sample.}
\label{fig:1}
\end{figure}
\begin{figure}
\centering
\includegraphics[width=0.4\textwidth]{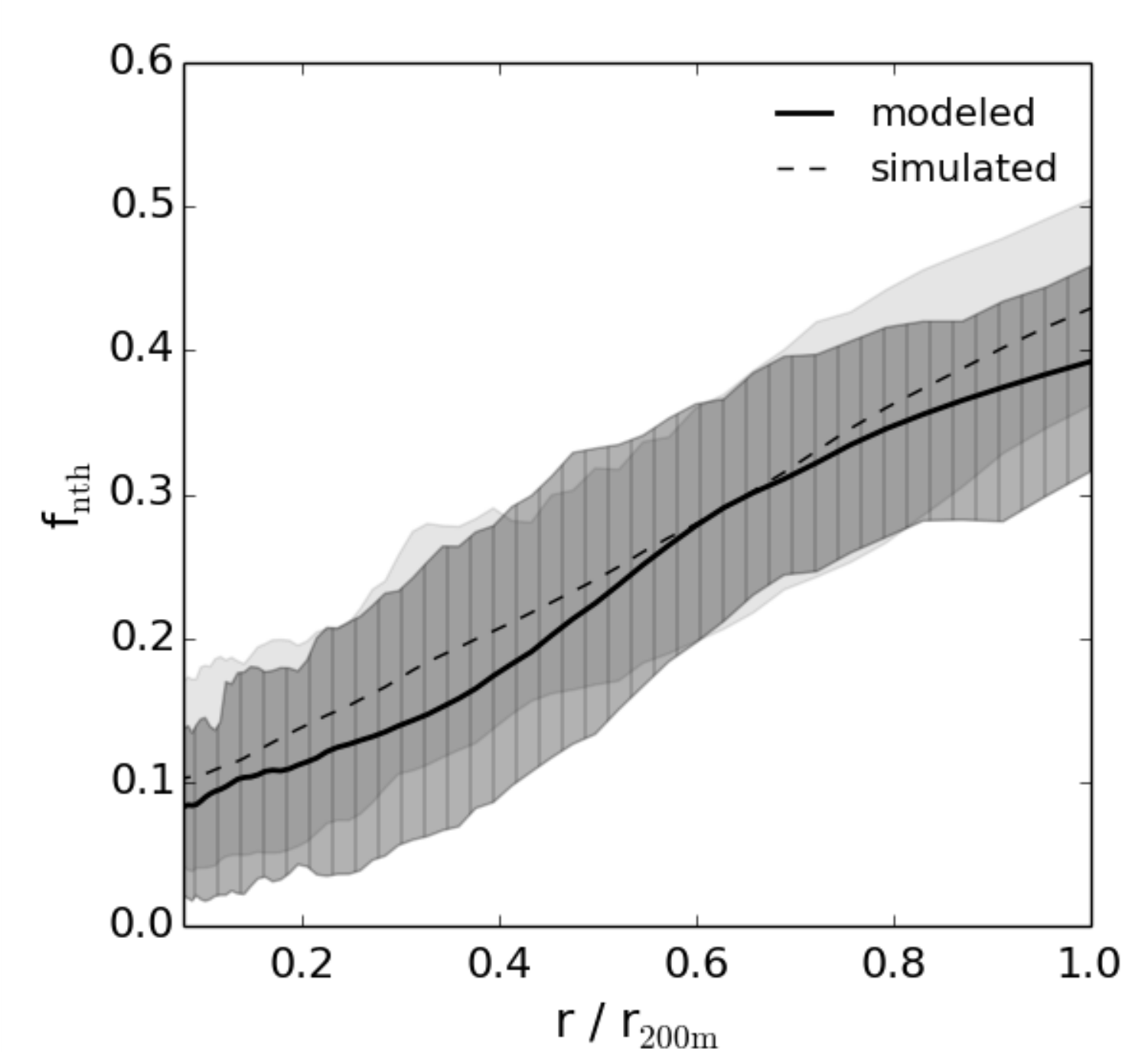}
\caption{Non-thermal fraction profile of the mass-limited sample. The solid
line and the hatched shaded region are the mean and the 16/84 percentile
of the modelled profiles; the dashed line and the unhatched shaded region are
those of the simulated profiles.} 
\label{fig:2}
\end{figure}
\begin{figure}
\centering
\includegraphics[width=0.45\textwidth]{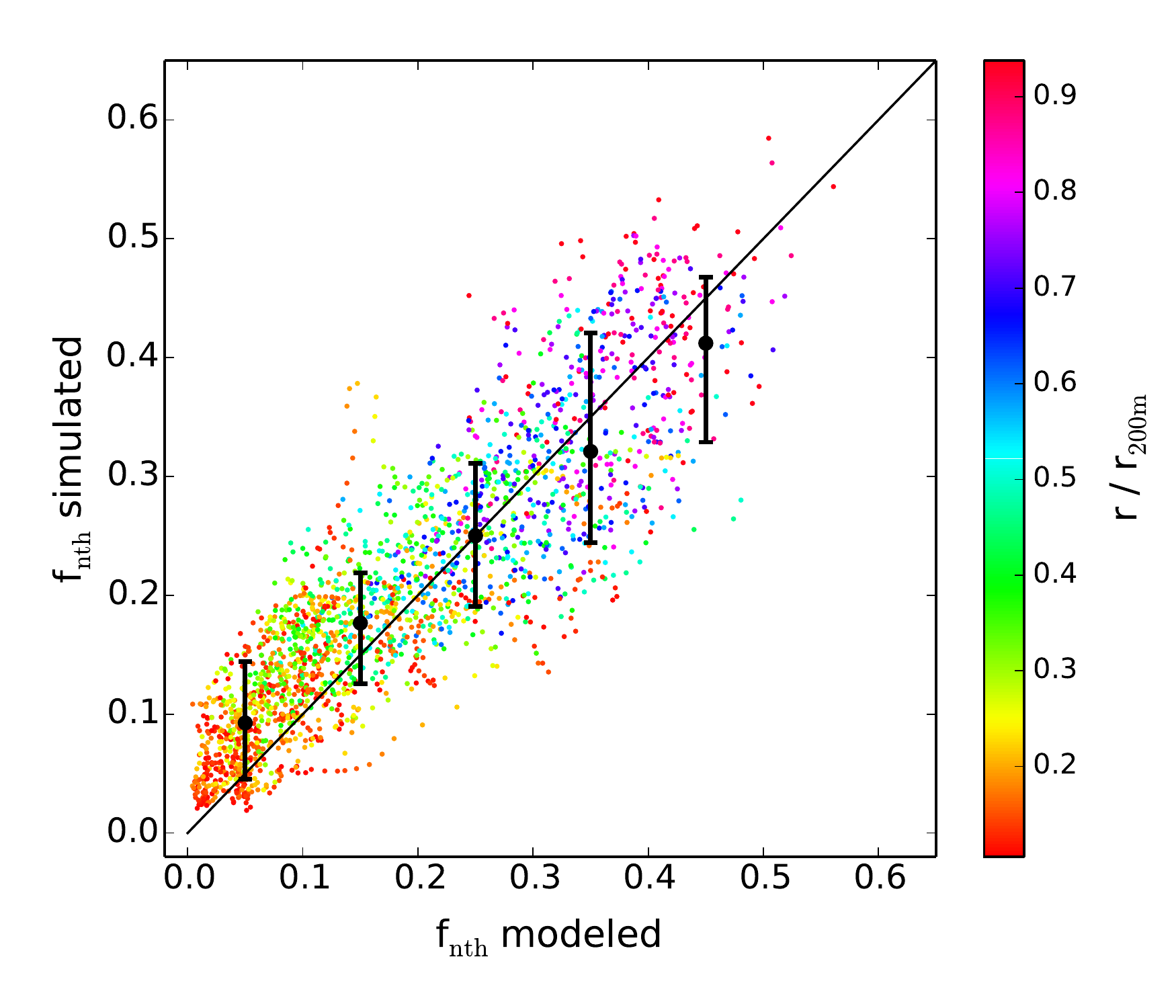}
\caption{Comparison of the modelled and simulated non-thermal fraction,
 $f_{\rm nth}$,  of the  mass-limited sample.
Each point on the scatter plot shows one radial bin of one cluster in
 the sample, and is colour-coded according to the central radius of the
 bin relative to $r_{200m}$. Only radial bins between $0.1$ and $1$ $r_{200m}$ are shown. The black points with error bars show the median and 16/84
percentile of the distribution of $f_{\rm nth}$ measured from the simulation in
bins of modelled $f_{\rm nth}$ values. The diagonal line
 shows the one-to-one correspondence.}
\label{fig:3}
\end{figure} 

We show the comparison of the modelled and simulated
non-thermal fraction profiles of $6$ clusters in
Fig.\;\ref{fig:1}. The clusters are selected such that
their masses spread over the full range.
For all clusters shown, there is a clear trend of $f_{\rm nth}$ increasing with
radius in the simulated profiles. This trend is a natural consequence of an increasing turbulence dissipation
time at larger radii, and is well reproduced by the modelled
profiles. 
On the other hand, the values of the non-thermal fraction at the same radius
scaled by $r_{\rm 200m}$ vary by a factor of a few among the clusters. This
distinctive difference in the $f_{\rm nth}$ values is also well
reproduced by the modelled profiles.

The mean $f_{\rm nth}$ profiles of the whole sample are shown
in Fig.\;\ref{fig:2}. The solid and the dashed lines are the modelled and
simulated profiles, respectively.  
Not only does the mean agree, but also the magnitude of the
scatter (shown by the shaded regions) agrees.

Fig.\;\ref{fig:3} shows a more quantitative comparison of the modelled and
simulated $f_{\rm nth}$ values. Each data point here shows the modelled versus
simulated $f_{\rm nth}$ values in one logarithmic radial bin of one cluster in
the sample. Larger $f_{\rm nth}$ values are found at larger radii, as shown by
the colour-coding. To guide the eye, we group the data points into bins according
to their modelled $f_{\rm nth}$ values, and mark the median simulated $f_{\rm nth}$ value of each group with a black point whose x-position indicates the center of the bin. The
associated error bar shows the $1\sigma$ scatter of the simulated $f_{\rm
nth}$ distribution. We find an excellent agreement between
the modelled and simulated $f_{\rm nth}$.

Looking closer, the slight deviation of the black points
from the one-to-one 
relation (the diagonal line) at large $f_{\rm nth}$ values can be explained by
the selection effect that only data points between $0.1$ and $1$ $r_{200m}$ are
shown. The same selection effect does not seem sufficient to explain the
deviation at small $f_{\rm nth}$ values, and this may suggest a
systematic tendency of a smaller modelled than simulated non-thermal
fraction at $r< 0.25 r_{\rm 200m}$. Although the statistical
significance is only $1\sigma$, we offer a
possible explanation of this deviation in Sect.\;\ref{sec:dis}. 

\subsection{Effect of varying model parameters}
\label{sec:varyparam}

The two parameters in the analytical model, $\eta$ and $\beta$, are physical
parameters related to turbulence injection and dissipation, respectively.
However, their values are not yet well-constrained from theory.
In SK14, we find that $\eta \beta \approx 0.7$ and $\beta \approx 1$
provide excellent agreement between the model predictions and
the fitting formulae derived from the existing observations \citep{arnaud10,
planck13a} and numerical simulations \citep{shaw10,bat12}. The same
values reproduce the simulation outputs used in this paper.

\begin{figure*}
\centering
\includegraphics[width=0.7\textwidth]{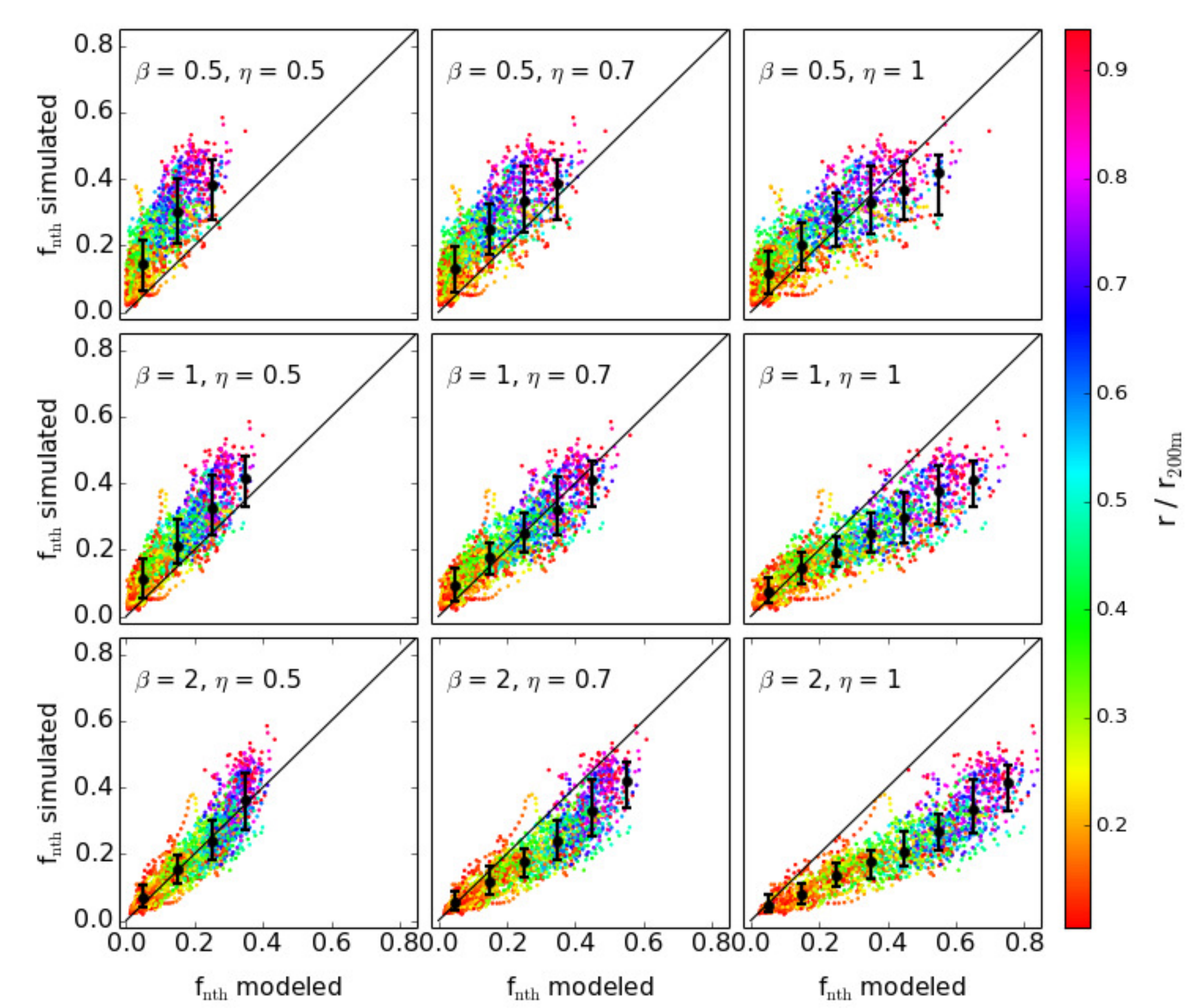} 
\caption{Effect of varying the parameters
$\beta$ and $\eta$. In each panel the symbols are the same as those in
Fig.\;\ref{fig:3}. The central panel with $\beta=1$ and
 $\eta=0.7$ is identical to Fig.\;\ref{fig:3}.}
\label{fig:4}
\end{figure*}

To examine how sensitive the 
comparison results are to the exact values of $\eta$ and
$\beta$, we show the effects of varying them in  Fig.\;\ref{fig:4}.
Each panel in Fig.\;\ref{fig:4} uses different values of
$\beta$ and $\eta$ as shown, and the central panel with $\eta=0.7$ and
$\beta=1$ is identical to Fig.\;\ref{fig:3}. 

When the cluster mass growth is fast,
i.e., when $\sigma^2_{\rm tot}$ increases with a timescale $t_{\rm
growth}$ shorter than the turbulence
dissipation time-scale $t_{\rm d}$, the non-thermal fraction approaches
$\eta$ (Sect.\;$3.2$ in SK14). In the opposite case, the non-thermal
fraction approaches $\eta t_{\rm d}/t_{\rm 
growth} \propto \eta\beta$. At $z \approx 0$, $t_{\rm d} \ll t_{\rm
growth}$ in the inner region of a galaxy cluster, whereas 
$t_{\rm d}/\beta$ is comparable to $t_{\rm growth}$ in the
outskirts. This suggests that 
$f_{\rm nth}$ is roughly proportional
to $\eta\beta$ when $\beta<1$, and the shape of the radial dependence of $f_{\rm
nth}$ is mainly given by the increase of the dynamical time with radius. For
larger values of $\beta$, the radial dependence of $f_{\rm
nth}$ should flatten towards large radii due to the saturation of $f_{\rm
nth}$ to the value of $\eta$ in the fast growth regime. 

These features are clearly visible in Fig.\;\ref{fig:4}: the slope of the
modelled versus simulated $f_{\rm nth}$ relation is primarily determined
by $\eta\beta$, and the curvature of the relation by $\beta$. As far as
the slope is concerned, the three panels on the diagonal
from bottom left to top right with $0.5 \leq \eta \beta \leq 1$ provide a
good match between the modelled and simulated values. From the curvature of the
relation, the central panel with the default parameter values give the best
agreement, in the sense that the scatter of the data points at each radius
(with each colour) is most symmetric around the one-to-one relation.

\subsection{Dynamical state}
\label{sec:dyn}

SK14 used the analytical mean mass accretion history of
\citet{zhao09} to show that the average non-thermal pressure fraction increases
with the cluster mass and redshift. This feature is hard to test
directly with the simulated cluster sample described in Sect.\;\ref{sec:sim}
due to the limited range of masses and redshifts for which the profiles
of the clusters are well-resolved. Also, as discovered by NLN14, the
redshift and mass dependences are greatly reduced when the cluster
radius is scaled by $r_{\rm 200m}$.

Still, we can divide the simulated cluster sample by their
accretion histories, and test whether the model and the simulation yield
the same difference on $f_{\rm nth}$ between the sub-samples.
Since the model attributes the origin of the mass and redshift
dependences of $f_{\rm nth}$ to the dependence on the recent mass
accretion history, this provides a more direct test of the model
prediction than comparing the average non-thermal pressure fraction of
cluster samples at different redshift or with different masses. 

\begin{figure}
\centering
\includegraphics[width=0.38\textwidth]{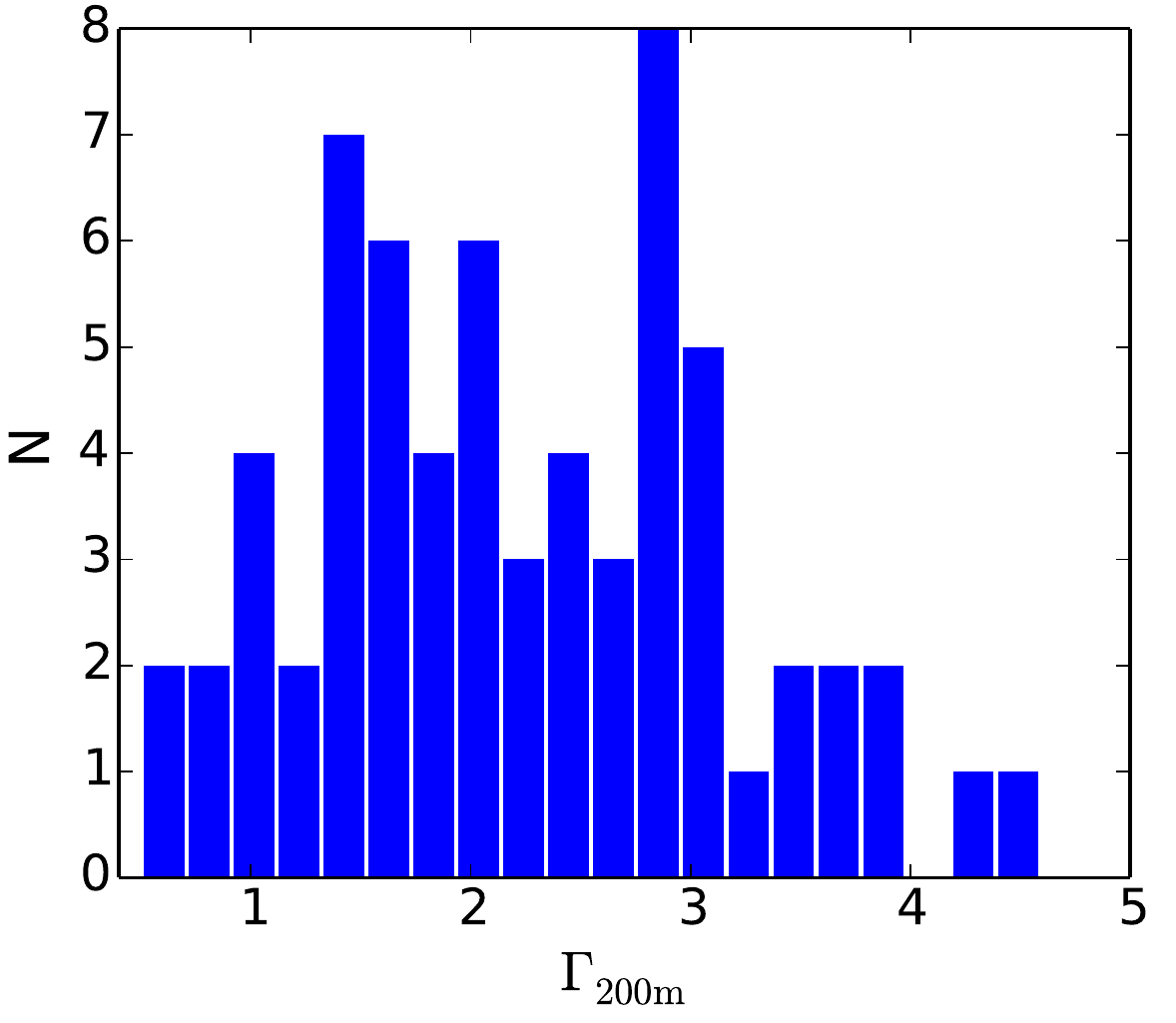}
\caption{Distribution of the proxy of the accretion history and dynamical state,
$\Gamma_{\rm 200m}$, computed from the mass-limited sample of simulated galaxy
clusters.
}
\label{fig:Ghist}
\end{figure}

\begin{figure}
\centering
\includegraphics[width=0.47\textwidth]{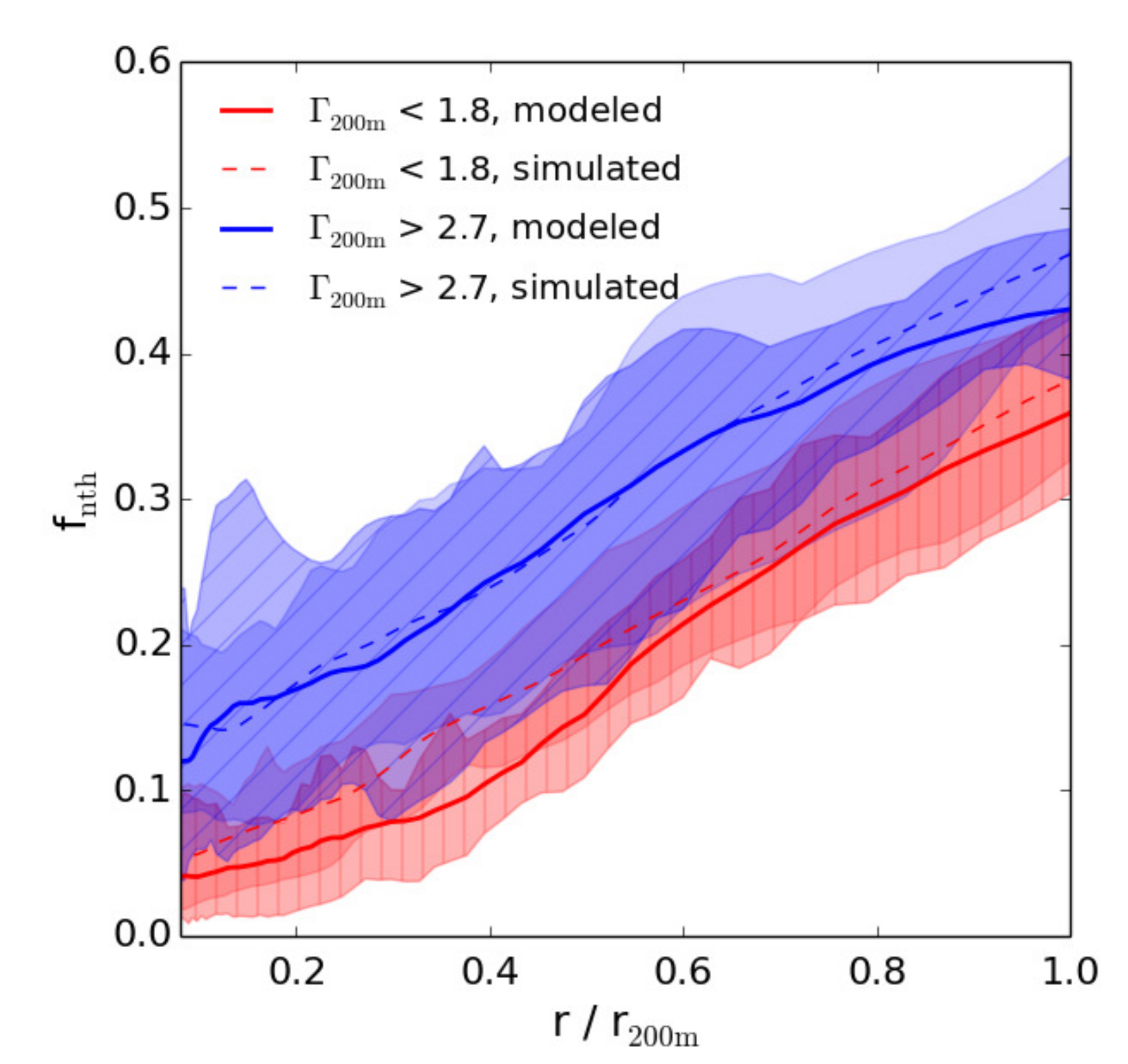}
\caption{modelled and simulated $f_{\rm nth}$ profiles of an early growth
sub-sample ($\Gamma_{\rm 200m}<1.8$, blue lines) and a late growth sub-sample
($\Gamma_{\rm 200m}>2.7$, red lines). The lines and the shaded regions are the mean profile
of the sample and the 16/84 percentiles, respectively.  }
\label{fig:7}
\end{figure}

We adopt a simple quantification of the recent accretion history as introduced
by NLN14 and \citet{diemer14},
\eq{
\Gamma_{\rm 200m} \equiv \frac{\log_{10}[M_{200m}(z=0)] - \log_{10}[M_{200m}(z=0.5)]}{
\log_{10}[a(z=0)] - \log_{10}[a(z=0.5)]}\,.
}
A larger $\Gamma_{\rm 200m}$ value indicates more mass growth since $z =
0.5$. The value of $\Gamma_{\rm 200m}$ is also an indicator of the 
dynamical state, as there is a strong correlation between
the recent mass growth and the dynamical state of a galaxy cluster, 

The distribution of $\Gamma_{\rm 200m}$ in the mass-limited cluster sample is
shown in Fig.\;\ref{fig:Ghist}. We select two sub-samples of the simulated clusters with
$\Gamma_{\rm 200m}<1.8$ and $\Gamma_{\rm 200m}>2.7$, respectively. Both
sub-samples contain $23$ galaxy clusters. We apply the analytical model to
each cluster in the sub-sample and compare the mean $f_{\rm nth}$ profile of
each sub-sample with the corresponding simulated one.
As shown in Fig.\;\ref{fig:7}, the sub-sample with higher recent mass growth
has a significantly higher non-thermal pressure fraction at all radii. This is
consistent with the previous numerical studies
\citep[e.g.,][]{nagai07b,piff08,nelson12} which consistently
find a larger hydrostatic mass bias for less relaxed, recently merged clusters. 
This difference in the average non-thermal pressure fraction is remarkably
 well reproduced by the analytical model. This result reinforces
 the basic underlying 
 physical picture that intracluster turbulence is triggered during
 the cluster mass assembly, 
 and that the kinetic energy in the intracluster turbulence is derived
 ultimately from the gravitational energy released during the structure growth. 


\begin{figure}
\centering
\includegraphics[width=0.42\textwidth]{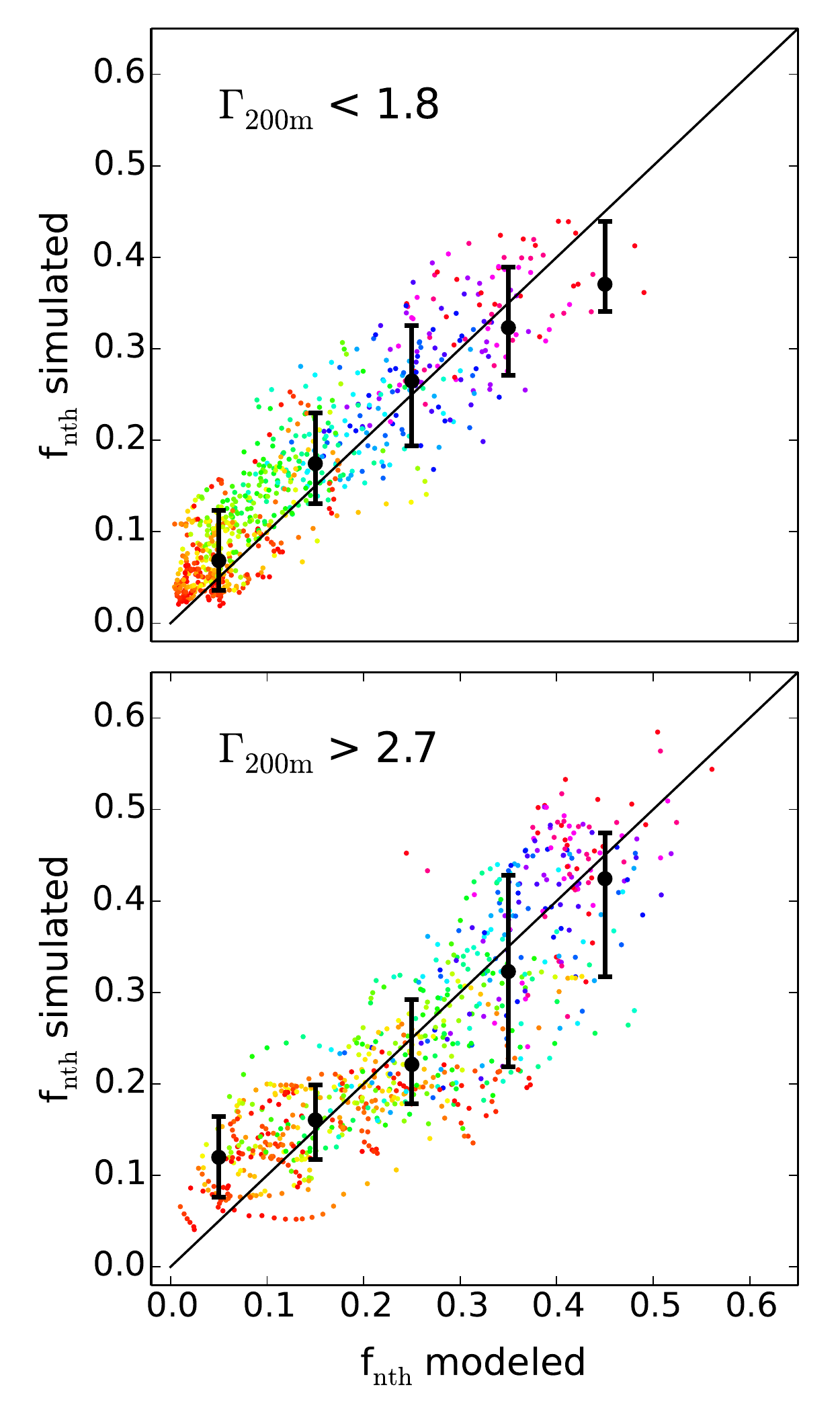}
\caption{Comparison of modelled and simulated $f_{\rm nth}$ of the early growth
sub-sample (upper panel) and the late growth sub-sample (lower panel). In each
panel the symbols are the same as those in Fig.\;\ref{fig:3}. }
\label{fig:6}
\end{figure}

In Fig.\;\ref{fig:6} we compare the modelled and simulated
non-thermal fractions in each radial bin of each cluster in the two sub-samples.
It is clear that, for the early accretion ($\Gamma_{\rm 200m}<1.8$) sub-sample which
consists of more dynamically relaxed clusters at the time of comparison
($z=0$), the scatter between the modelled and simulated $f_{\rm nth}$
values is smaller. This is in 
accord with the expectation that the analytical model works better for
dynamically relaxed clusters. Nevertheless, a clear correlation exists also for
the more disturbed clusters ($\Gamma_{\rm 200m}>2.7$), suggesting that the analytical
model is also applicable to these systems in estimating the turbulence pressure,
though with greater noise.

We note that the $\Gamma_{\rm 200m}$ parameter used in this paper is not optimal
as a proxy for the dynamical state, since it is defined with the mass increase
between two snapshots. By definition, the dynamical state of a cluster can be determined
by its temporary state. A dynamical state proxy defined at a single snapshot based
on the dynamical properties of the halo particles would be more convenient to
use, and at the same time provide a more direct characterization of the
deviation from hydrostatic equilibrium. For future studies of assigning
the non-thermal pressure profile to dark matter haloes extracted from the
dark-matter only N-body simulation, such more advanced dynamical state
proxy may be preferred.

%%%%%%%%%%%%%%%%%%%%%%      sec       %%%%%%%%%%%%%%%%%%%%%%%%%
\section[]{Discussion: test of the physical paradigm}
\label{sec:dis}

Evolution of the intracluster turbulence is a problem involving a vast range of
spatial- and time-scales. The relevant physical processes include the 
cluster mass assembly in a cosmological context, the merger and
accretion shocks which convert the bulk kinetic energy into the
turbulence kinetic energy and heat, and the detailed intracluster gas
dynamics associated with the development and cascade of turbulence. 
Simulating all of them with sufficient numerical precision is beyond the
reach of a single set of numerical simulations. Simulations dedicated to certain
physical processes would be needed for testing them in greater detail.

In this respect, the large-size cosmological simulation used in this paper is
ideal for testing the relation of turbulence growth with cluster mass assembly in a
cosmological context, for which the picture underlying the analytical model
has been verified by the positive results presented in Sect.\;\ref{sec:nonth}.
On the other hand, cosmological simulations of a single cluster
\citep[e.g.,][]{vazza09,vazza11,paul11,min14} are better suited 
for studying mechanisms of turbulence injection, and high resolution
simulations performed on a fixed grid \citep{gas13, gas14} are better suited
for studying the turbulence cascade process. The insights gained from
these dedicated simulations can be easily incorporated into the
analytical model.

For a precise assessment of the amplitude of intracluster turbulence pressure,
it is important to know the effective thermalization ratio at turbulence
injection, and the turbulence dissipation time-scale. In the
framework of the SK14 analytical model, this suggests the need to
determine the values of the model parameters $\eta$ and $\beta$ and investigate
their possible dependence on radius, redshift, and cluster mass. These can in
principle be realized by dedicated numerical simulations, when numerical
effects in these simulations are well-understood and controlled. Recently,
using the moving-mesh numerical scheme, \citet{schaal15} reported a higher
energy dissipation fraction contributed by shocks in the warm hot intergalactic
medium, and correspondingly a higher average Mach number of shocks at which the
bulk of energy dissipates, than previous studies performed with the Adaptive
Mesh Refinement technique \citep[e.g.,][]{ryu03,pfr06,kang07,vazza11,pla13}.
This, if confirmed, would suggest a higher thermalization ratio, and that a
radius and redshift dependence of $\eta$ would be determined by the relative
importance of the high Mach number accretion shocks and the low Mach number
internal shocks.

We note that the SK14 analytical model is not consistent with the long-term
power law decay behaviour expected for the turbulence kinetic energy
\citep{landau59,frisch95,sub06}. This inconsistency is due
to our assumption of a one-to-one relation between the cluster radius and the
turbulence dissipation time-scale, which is the ratio of the
size and velocity of the largest eddies, at that radius (i.e., $t_{\rm
d}\propto t_{\rm dyn}$).  The consequence of this assumption is
most visible in the regions where turbulence dissipates 
much faster than it grows, and may have contributed to the possible systematical
difference between the modelled and simulated $f_{\rm nth}$ at small radii. To
correct for this, one may need to include a spectral dimension to the model,
that is to keep track of the power spectrum of turbulence velocity field at each
radius as a function of time. This, in turn, will allow for an easier link to
intracluster magnetic fields and cosmic rays.

%%%%%%%%%%%%%%%%%%%%%%      sec       %%%%%%%%%%%%%%%%%%%%%%%%%
\section[]{Conclusion and perspectives}
\label{sec:con}

We have compared the SK14 analytical model for the turbulence pressure inside
galaxy clusters to a state-of-the-art hydrodynamics numerical
simulation. 
The analytical model and the simulation outputs show
excellent agreement on the non-thermal 
pressure fraction on a cluster-by-cluster basis
- both its radial profile and its dependence on the cluster mass
accretion history.   

This demonstrates that the SK14 model in its current form can already be used
to predict the amplitude of intracluster turbulence pressure with a precision
comparable to that of the state-of-art cosmological hydrodynamics simulations.
This opens up an exciting possibility that we may be able to
use the analytical model to correct the systematic bias in the mass
estimation of galaxy clusters due to the turbulence pressure. The analytical
model, in turn, would also provide a convenient and efficient way to
interpret the SZ power spectrum and observations of cluster outskirts from ongoing and upcoming large cluster surveys. 

At the same time, the comparison results show that a simple analytical
model can indeed capture the basic physical processes related to the evolution of intracluster turbulence pressure.
In particular, our comparison study has verified the underlying
physical picture that the turbulence growth is determined by
cluster mass assembly in a cosmological context. The detailed physics
regarding injection 
and dissipation of intracluster turbulence requires further tests from
comparisons with dedicated high resolution simulations of individual clusters.
We point out that adding a spectral dimension to the model may lead to a better description of the
long-term dissipation of the turbulence, further improve the consistency
with simulations in the inner regions of clusters, and provide a framework for
a unified understanding of non-thermal phenomena in galaxy clusters.

\section*{Acknowledgements}
XS thanks Massimo Gaspari for helpful discussions.
KL and DN acknowledge support from NSF grant AST-1009811, NASA ATP grant
NNX11AE07G, NASA \textsl{Chandra} grants GO213004B and TM4-15007X, and the
Research Corporation. This work was supported in part by the facilities and staff of
the Yale University Faculty of Arts and Sciences High Performance Computing
Center. We thank Erwin Lau and Peng Oh for comments on the draft.

\bibliographystyle{mn2e}

\appendix

\section[]{Measure gas velocity dispersion from simulation}
\label{sec:ap1}

\begin{figure}
\centering
\includegraphics[width=0.45\textwidth]{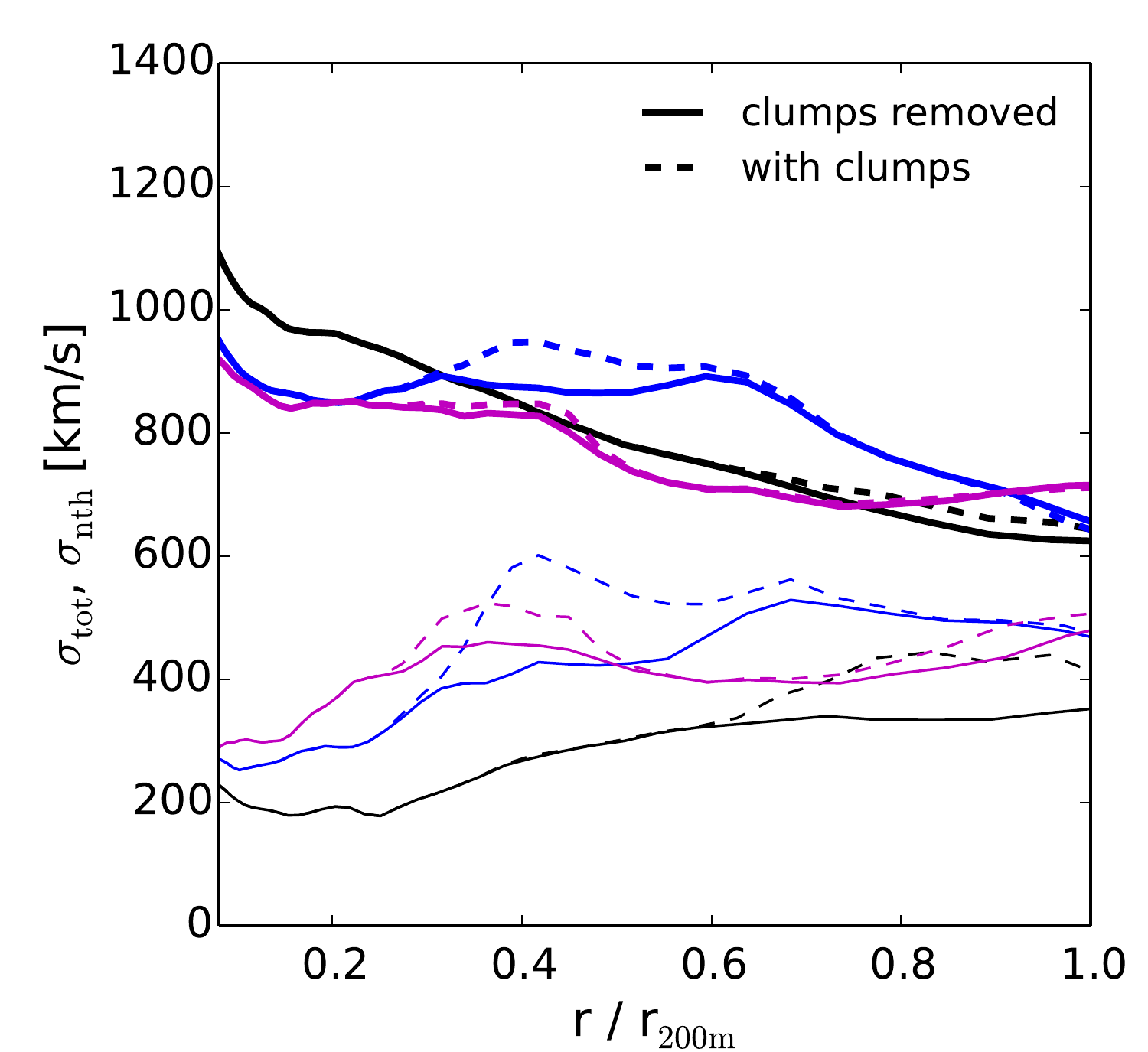}
\caption{ Measured $\sigma_{\rm tot}$ (thick lines) and $\sigma_{\rm
nth}$ (thin lines) profiles with and without removing the sub-structures (clumps) in the
simulation. 
Profiles for three clusters at $z=0$ with various mass accretion histories are
presented (black: early accretion; blue: medium; magenta: late accretion). All
three chosen clusters have masses close to the median mass of the sample. }
\label{fig:8}
\end{figure}

Following NLN14, we measure the non-thermal velocity dispersion of
the gas $\sigma_{\rm nth}$ in radial shells after subtracting the mean velocity
of the shell with respect to the centre-of-mass velocity of the total mass
interior to this radial shell.
In order
to remove the kinetic energy associated with sub-structures which does not contribute to the
pressure of the global intracluster gas, we also exclude the contribution from
gas that lies in the high-density tail in the probability distribution
of gas densities according to the
procedure presented in \citet{zhu13b}. 
 We show the effect of removing the sub-structures to the
$\sigma_{\rm tot}$ and $\sigma_{\rm nth}$ profiles in Fig.\;\ref{fig:8}. The
sub-structures generally affect $\sigma_{\rm nth}$ more than $\sigma_{\rm tot}$.
Without removing them, the non-thermal fraction measured from simulations
would be slightly over-estimated. No strong trend of sub-structure influence
with accretion history is observed.
In addition to sub-structure removing, we smooth the profiles with the
Savitzky-Golay filter used in \citet{lau09}.

 We note that there are different choices of the mean velocity
to be subtracted from the velocity field when computing non-thermal velocity
dispersion from simulations. They correspond to different ways
of decomposing the velocity field. Studies focusing on the
turbulence properties in the inertial range usually adopt the averaged velocity in a local volume as the
mean velocity \citep[see e.g.][]{vazza12}. This decomposes the velocity field
into parts that are smaller or larger compared to the size of the chosen local
volume, which are usually referred to as `turbulence' and `bulk motion',
respectively.  The spherical averaging method we use decomposes the
velocity field into the average infall/outflow motion and the residual motions.
These residual motions, which receive contribution from both turbulent random
motion and some large scale bulk motions, are the main source of the HSE mass
bias. Further studies are required to gain more understanding of the nature of gas motions in the cluster outskirts. This would help distinguish the
physical sources of the non-thermal velocity dispersions, and point to an optimal way of
decomposing the velocity field that is conceptually clear and at the same time
matches the methods used in analysing observations. For the moment, we stick to
the spherical averaging method. The $\sigma_{\rm nth}$ measured this way is
clearly defined, and its contribution to the hydrostatic mass bias is
relatively well-understood \citep{lau13}.

The velocity dispersions we measure here are averaged over all directions,
but physically only the radial velocity dispersion contributes to the
pressure support against gravity. 
Shown in \citet{lau09,nelson12} and NLN14, the gas motions are
predominantly radial at cluster outskirts, especially near $r_{\rm 200m}$. The
physical origin of the measured velocity anisotropy is not yet clear.
Therefore, velocity anisotropy is so far dismissed in the analytical model and the comparison to simulations. Once the amplitude of
velocity anisotropy and its radial dependence are known, it can be easily
taken into account.

\section[]{Visualizing propagation of merger shock}
\label{sec:ap2}

\begin{figure}
\centering
\includegraphics[width=0.43\textwidth]{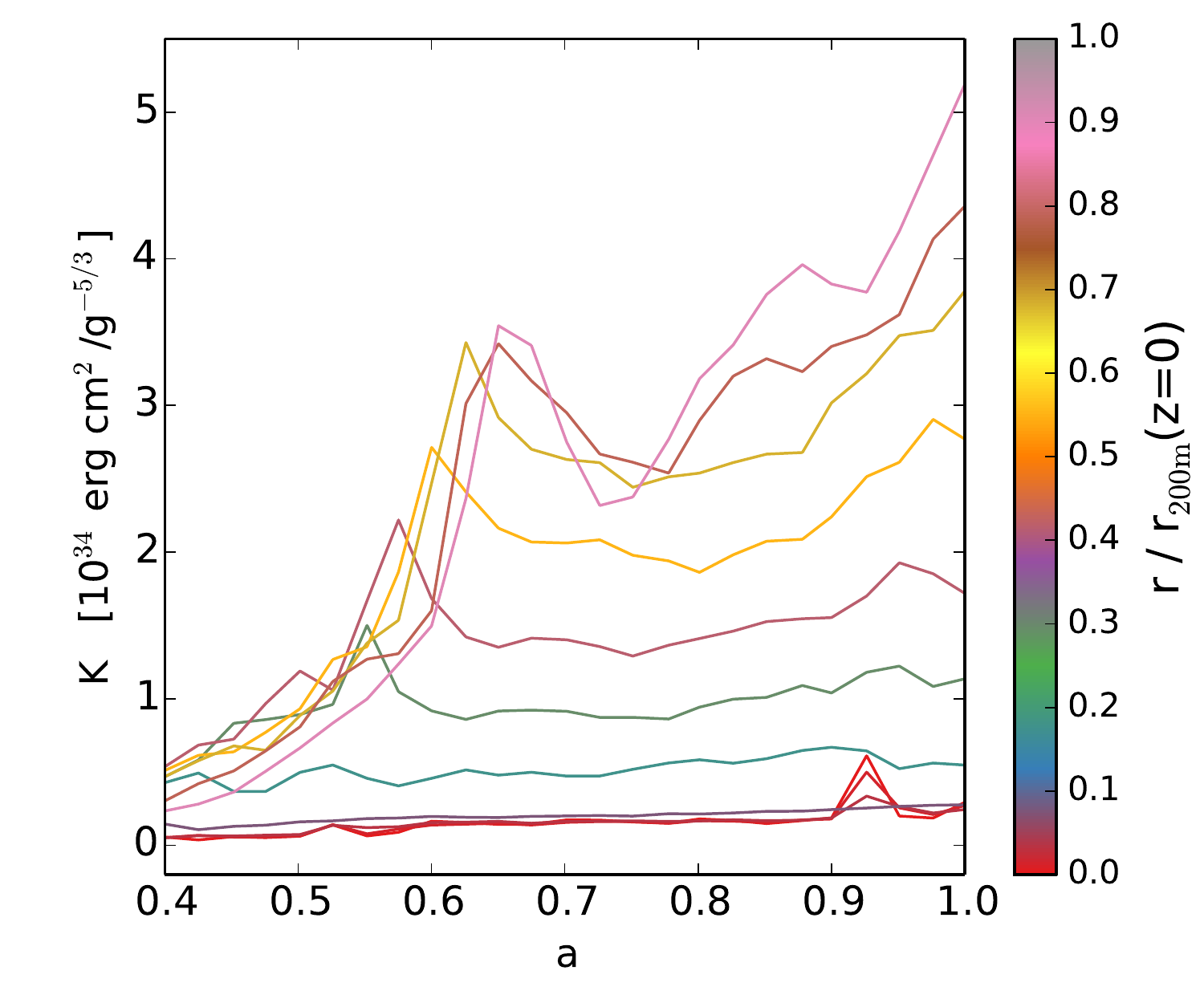}
\caption{Evolution of the average entropy $K \equiv T/\rho_{\rm
gas}^{2/3}$ in Eulerian shells for the same simulated cluster used for
Fig.\;\ref{fig:02}, where the mass-weighted temperature $T$ and the gas
density $\rho_{\rm gas}$ are averaged in the shell.
}
\label{fig:entropy_jump}
\end{figure}

\begin{figure*}
\centering
\includegraphics[width=1\textwidth]{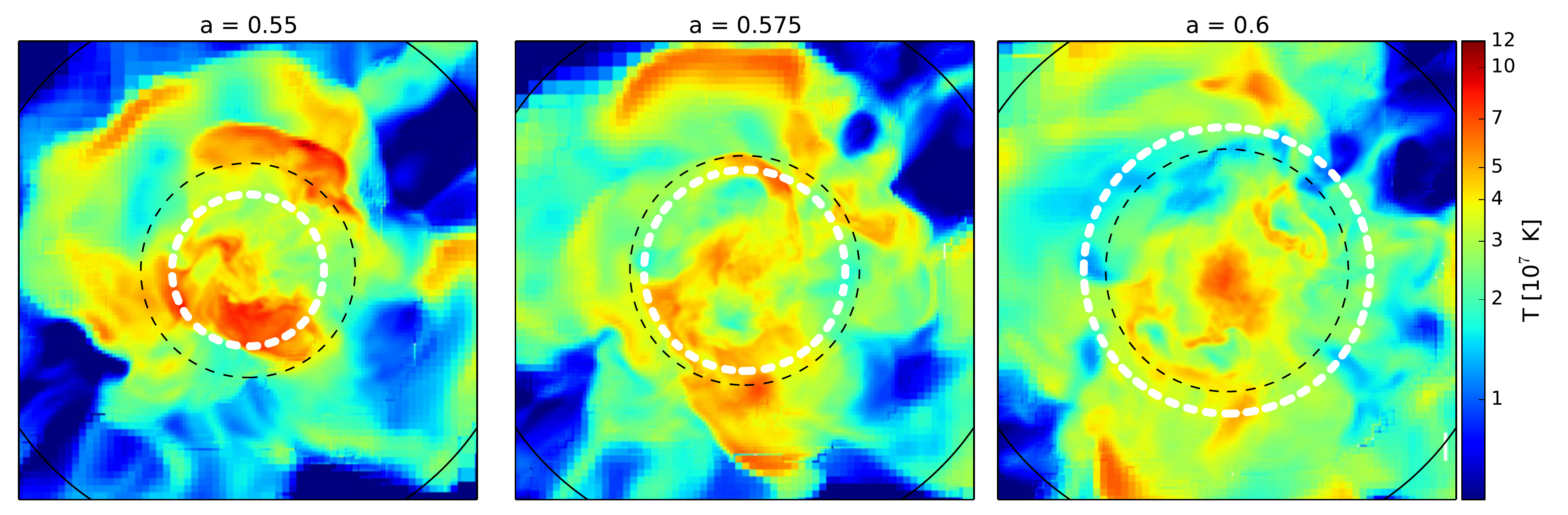}
\caption{Gas temperature distribution of the simulated cluster used for
Figs.\;\ref{fig:02} and \ref{fig:entropy_jump} at three consecutive
snapshots. The size of the images is 4\;Mpc/h, and the mass-weighted temperature
is averaged in a slice of thickness $\simeq$ 500\;kpc/h across the cluster
centre. The white dashed circles mark the radial location of the peak in the
$\sigma_{\rm tot}$ profile that corresponds to the `wiggles' in
Fig.\;\ref{fig:02} between $a=0.5$ and $0.7$. Note that the radial extension of the `wiggles' is large, as can be seen
from Fig.\;\ref{fig:02}. The black dashed and solid circles show the position of
$r_{\rm 200m}$ at the time of the snapshot and at $z=0$ respectively.}
\label{fig:T_shock}
\end{figure*}
 
While inspecting the evolution of $\sigma_{\rm tot}$ in the simulations at
fixed Eulerian radii, we discover some `wiggles' - sharp rise and fall in
$\sigma_{\rm tot}$ with time (Fig.\;\ref{fig:02}). In
Fig.\;\ref{fig:entropy_jump} we show that there are indeed entropy jumps at the positions of these `wiggles',
supporting that they originate from merger shocks. The magnitude of the jumps
is typically $\lesssim 2$, corresponding to low Mach numbers that are expected
for the merger shocks. Note that the entropy in one shell also falls after the
shock, likely due to the expansion of the shock-heated gas into a neighbouring
shell at a larger radius.

These low Mach number merger shocks are close to sound waves - they are more
efficient in compressing the gas than increasing its entropy. Adiabatic
compression contributes to most of the temperature jump at these shocks.
As a consequence, the temperature experiences an evident jump at the shock, and
falls close to its original value after the shock passes. The temperature map,
therefore, enables one to see the propagation of the shock. In
Fig.\;\ref{fig:T_shock}, one sees spatially coherent high temperature regions
elongated in the azimuthal direction, indicative of shock fronts. The expansion
of the shock fronts inside the cluster coincide with the increase of peak radius
of the corresponding `wiggles' in Figs.\;\ref{fig:02} and
\ref{fig:entropy_jump} (white dashed circles).

In one-dimension, due to the projection effect and the possible intrinsic asymmetry of shock
fronts in different azimuthal directions, shocks are much more prominent as
`wiggles' in temperature or $\sigma_{\rm tot}$ against time rather than against
radius.
Thus plots like Figs.\;\ref{fig:02} and \ref{fig:entropy_jump} can better
depict the propagation of shocks than profiles of thermodynamical quantities.

\label{lastpage}
\end{document}